# Fast and Robust $T_1$ Mapping Based on a 3D Dual-Echo UTE Sequence (PETALUTE) for SPION Biodistribution Assessment

## Author

Zhen Jiang[1], Stephen Sawiak[2,3], Alexandra Lipka[1,5,6], Xin Shen[7], Uzay Emir[8], Ali Özen[9], Mark Chiew[10,11,12], Justin Geise[1], Joseph Speth[1], Deng-Yuan Chang[1], Jessica Veenstra[1], Mitchell Gabalski[13], Luis Solorio[13], Gregory Tamer Jr.[13], Matthew Scarpelli[1]

[1]School of Health Sciences, College of Health and Human Sciences, Purdue University, West Lafayette, IN, United States, [2]Department of Physiology, Development, and Neuroscience, University of Cambridge, Cambridge, United Kingdom, [3]Department of Clinical Neurosciences, Wolfson Brain Imaging Centre, Cambridge, United Kingdom, [5]College of Engineering, Purdue University, West Lafayette, IN, United States, [6]Division of Medical Physics in Radiology, German Cancer Research Center (DKFZ), Heidelberg, Germany, [7]Department of Radiology, University of California San Diego, San Diego, CA, United States, [8]Department of Radiology, University of North Carolina at Chapel Hill, Chapel Hill, NC, United States, [9]Department of Radiology, Medical Physics, Medical Center-University of Freiburg, Faculty of Medicine, University of Freiburg, Freiburg, Germany, [10]Nuffield Department of Clinical Neurosciences, University of Oxford, Oxford, United Kingdom, [11]Physical Sciences Platform, Sunnybrook Research Institute, Toronto, ON, Canada, [12]Department of Medical Biophysics, University of Toronto, Toronto, ON, Canada, [13]Weldon School of Biomedical Engineering, Purdue University, West Lafayette, IN, United States

## Abstract

### Background

Superparamagnetic iron oxide nanoparticles (SPIONs), such as ferumoxytol, are promising theranostic agents that can be assessed with MRI. Relaxation time mapping can provide reproducible and quantitative biomarkers of SPION distribution, suitable for longitudinal and cross-individual studies. However, conventional approaches suffer from strong susceptibility artifacts, long echo times (TE), and prolonged scan times, which limit the accurate quantification of SPION biodistribution.

### Purpose

To address the limitations of conventional approaches, this study aimed to develop a fast, $B_1$-corrected $T_1$ mapping protocol based on PETALUTE (a 3D dual-echo ultrashort echo time MRI sequence with rosette k-space trajectory) using a variable flip-angle (VFA) acquisition for quantitative mapping of ferumoxytol distribution.

## Method

Agarose phantoms containing 0-5000 ppm ferumoxytol were scanned on a preclinical 7T MRI system using PETALUTE and a vendor-supplied $T_1$ mapping (RARE-VTR). PETALUTE $T_1$ maps were computed from two VFA acquisitions (4° and 20°). Mean $R_1$ values were correlated with ferumoxytol concentration to assess the method's reliability. For in vivo feasibility testing, mice bearing 4T1 mammary tumors and flank tumors were assigned to the control (n = 1) or ferumoxytol-injected group (n = 2; 40 mg/kg i.v.). Abdominal MRI scans were performed 24 hours post-injection with both PETALUTE and RARE-VTR. Regions of interest in the thigh muscle, mammary tumors, and flank tumors were analyzed to compare the estimated $T_1$ and $R_1$ values obtained with both methods.

## Results

In the phantom study, PETALUTE showed positive contrast for ferumoxytol at all concentrations except the highest concentration (5000 ppm), whereas the conventional RARE-VTR could not produce a positive contrast. For PETALUTE, there was a significant linear correlation between $R_1$ and ferumoxytol concentration (R = 0.975, p < 0.01), while RARE-VTR showed no significant correlation (R = 0.672, p = 0.144). In vivo, PETALUTE provided high-resolution, non-gated, whole-abdominal images with short acquisition times (4 minutes 19 seconds). In ferumoxytol-injected mice, flank tumors exhibited $T_1$ shortening, consistent with the expected iron accumulation. The dual-echo capabilities of PETALUTE facilitated observation of elevated $T_1$ with preserved $T_2^*$-weighted signal in one of the mammary tumors.

## Conclusions

The proposed PETALUTE-based $T_1$ mapping enables fast, quantitative, and positive-contrast ferumoxytol imaging with higher spatial coverage and more stable measurements across a wider concentration range than conventional RARE-VTR $T_1$ mapping.

## Introduction

Nanoparticles (NPs) are a powerful platform for precision cancer therapy, enhancing the therapeutic index by concentrating agents in tumors while minimizing off-target exposure[1, 2]. In practice, NP uptake varies widely across patients and tumor types[3,4]. This variability makes the effective dose unpredictable, complicating treatment planning and clinical translation. Theranostic NPs, which combine therapeutic and imaging functions in a single platform, have therefore been used for the simultaneous treatment and visualization of NP delivery[5, 6, 7].

However, a noninvasive imaging modality is still required to confirm target engagement and enable personalized, response-adaptive dosing strategies[4].

Magnetic resonance imaging (MRI) is well-suited for theranostic NP tracking. It is non-ionizing, widely available, and offers high spatial resolution with excellent soft-tissue contrast. Superparamagnetic iron oxide nanoparticles (SPIONs), such as ferumoxytol, are MRI-visible theranostic NPs owing to their intrinsic physical properties[8]. They shorten both the transverse ($T_2/T_2^*$) and longitudinal relaxation times ($T_1$)[9, 10, 11], appearing brighter on $T_1$-weighted MRI images and darker on $T_2/T_2^*$-weighted MRI images.

However, MRI signal intensity alone is not a reliable quantitative indicator of SPION, as it is inherently arbitrary. Quantitative MRI (qMRI) is a more reliable approach that offers reproducible measurements grounded in physics models of the MR signal, derived from multiple images acquired with varied parameters to capture tissue-specific properties[12]. In particular, $T_1$ and $T_2/T_2^*$ mapping yield voxel-wise parametric maps that encode quantitative relaxation values, reflecting tissue composition and the local magnetic environment[13, 14, 15, 16]. These relaxation mapping techniques have also been investigated for assessing SPION biodistribution[9, 17, 18, 19, 20, 21], although several challenges remain in their application to SPION imaging.

$T_2/T_2^*$ mapping is sensitive to SPIONs, but its accuracy and efficiency decline as concentration increases[9]. At high concentrations, strong susceptibility artifacts cause the blooming effect and signal voids, degrading spatial localization and quantitative assessment[20]. In contrast, $T_1$ mapping is more robust to susceptibility artifacts and can offer a higher signal-to-noise ratio (SNR) than $T_2$ mapping[20], but it still has essential limitations in SPION imaging. First, conventional $T_1$ mappings, such as Inversion Recovery Spin-Echo (IR-SE) and Rapid Acquisition with Relaxation Enhancement with Variable TR (RARE-VTR), utilize echo times (TE) longer than the ultrashort $T_2$ of SPION-rich regions, resulting in $T_2$-related signal loss and inaccurate $T_1$ calculations. In addition, a long repetition time (TR) is usually required to achieve complete relaxation for $T_1$ mapping, increasing susceptibility to motion artifacts and hindering clinical feasibility.

Ultra-short echo time (UTE) MRI is a promising technique to address the challenges described above[21, 22]. It can capture rapidly decaying signals and provide positive $T_1$ contrast for SPIONs[23]. However, the UTE MRI signal can be challenging to interpret in vivo because the short-$T_2^*$ signal from SPION is often masked by the long-$T_2^*$ signal from water and fat (e.g., necrotic tissue)[22, 24] or confounded by other short-$T_2^*$ components such as calcification or hemorrhagic byproducts[25]. Dual-echo UTE methods have been developed to separate short- and long-$T_2^*$ components, but, to our knowledge, their applications have mostly been tissue-specific rather than focused on SPION tracking[26, 27]. A dual-echo UTE $T_1$ mapping approach needs to be adapted for SPION monitoring.

To achieve this, we propose a $T_1$ mapping protocol based on PETALUTE, a 3D dual-echo UTE MRI sequence with a rosette k-space trajectory[28, 29]. PETALUTE acquires two echoes with ultrashort and longer TEs in a single acquisition. The dual-echo design, combined with an efficient sampling pattern that minimizes coherent undersampling artifacts, is particularly advantageous for SPION imaging. Furthermore, PETALUTE offers 3D high spatial resolution with a short acquisition time, which facilitates its clinical translation.

In this study, we developed and evaluated the performance of the PETALUTE $T_1$ mapping protocol for ferumoxytol quantification in vitro and in vivo. By combining PETALUTE with variable-flip-angle (VFA) acquisition[30] and $B_1$ map correction[31], we generated quantitative $T_1$ maps for ferumoxytol assessment. We hypothesize that PETALUTE will maintain a positive contrast of ferumoxytol while reducing susceptibility artifacts, even at high concentrations, and will enable acquisition of two contrasts with different TEs. To test this hypothesis, we compared PETALUTE $T_1$ maps with conventional RARE-VTR $T_1$ maps across a range of ferumoxytol concentrations in phantoms. In addition, we assessed the in vivo feasibility of using PETALUTE $T_1$ maps for evaluating ferumoxytol in a dual-tumor mouse model.

## Materials and Methods

### $T_1$ Map Calculation

PETALUTE $T_1$ mapping workflow is illustrated in **Figure 1**. We first generated apparent $T_1$ maps using PETALUTE first echoes from the VFA acquisitions (4°, i.e., the Ernst angle, and 20°) using **Equation 1**, based on the spoiled gradient-echo signal equation[30, 32].

$$T_{1,app} = 2\,TR\, \frac{S_1/\alpha_1 - S_2/\alpha_2}{S_2 \cdot \alpha_2 - S_1 \cdot \alpha_1} \qquad \text{Equation 1}$$

$T_{1,\,app}$, represents the apparent $T_1$ value; and $S_1$, $S_2$ are signal intensities from PETALUTE first echoes acquired with flip angles of 4° ($\alpha_1$) and 20° ($\alpha_2$), respectively.

Under a fully relaxed condition (TR ≥5$T_1$), the ratio between the two signal intensities, r, can be written as[31]

$$r = S_{1,\infty}/S_{2,\infty} = sin\alpha_1/sin\alpha_2 \qquad \text{Equation 2}$$

The PETALUTE sequence, however, uses a very short TR that prevents complete longitudinal relaxation. Therefore, a short-TR correction was applied to estimate the fully relaxed signals[33].

The actual flip-angle $\alpha_{1,\,act}$ was then calculated by combining **Equation 2** with the trigonometric multi-angle formula:

$$\alpha_{1,act} = \sqrt{5 + \frac{4}{r}} \qquad \text{Equation 3}$$

The corresponding voxel-wise $B_1$ map was then computed as the ratio of the actual flip-angle to nominal flip-angle (**Equation 4**).

$$B_1(x,y,z) = \alpha_{1,act}(x,y,z)/\alpha_1 \qquad \text{Equation 4}$$

Finally, the $B_1$ map was applied to correct the apparent $T_1$ map, producing the PETALUTE $T_1$ map (**Equation 5**).

$$T_1(x,y,z) = T_{1,app}(x,y,z) \cdot b1^{-2}(x,y,z) \qquad \text{Equation 5}$$

The RARE-VTR $T_1$ maps were computed by voxel-wise nonlinear least-squares fitting of a mono-exponential recovery model **(Equation 6)** to the multi-TR data.

$$S(TR) = S_0(1 - e^{-TR/T_1}) \qquad \text{Equation 6}$$

### Imaging Protocol and Reconstruction

Imaging protocol included vendor-provided RARE-VTR, Multi-Echo Gradient-Echo (MGE) and the proposed PETALUTE sequences. Parameters are summarized in **Table 1**. All scans were performed on a 7T horizontal-bore small animal MRI system (BioSpec 70/30; Bruker Instruments; maximum gradient strength: 660 mT/m; maximum slew rate: 4570 T/m/s).

RARE-VTR and MGE images were reconstructed on-scanner automatically using Bruker ParaVision 6.0.1. PETALUTE data were processed in MATLAB (202Xa, MathWorks, USA). The nonuniform fast Fourier transform (NUFFT) method was used to reconstruct the PETALUTE images using BART (Berkeley Advanced Reconstruction Toolbox; nufft for NUFFT reconstruction, ecalib for ESPIRiT coil-sensitivity estimation, and pics for iterative CS-SENSE) and MIRT (Michigan Image Reconstruction Toolbox; Gmri and ir_mri_density_comp to define the non-Cartesian encoding operator and density-compensation weights).

## Phantom Study

A 1 mL 0 ppm phantom (4.2% agarose) and six ferumoxytol (Feraheme, AMAG Pharmaceuticals) phantoms at targeted concentrations of 100, 250, 750, 1000, 1250, and 5000 ppm were prepared. Ferumoxytol was mixed into 4.2% agarose under heating and continuous stirring. Concentrations were verified by X-ray fluorescence. The ferumoxytol-agarose mixture was allowed to cool at room temperature until it gelled. A piece of the mixture was then scooped out and coated with a thin glue to keep it intact. Finally, the piece was embedded in a 1 mL microcentrifuge tube pre-filled with 4.2% agarose to form the ferumoxytol phantom. Embedding the small piece of ferumoxytol-infused agarose within 4.2% agarose created a compact phantom in which the ferumoxytol exhibits positive contrast relative to the surrounding agarose and remains compatible with the size constraints of the MRI coil. After cooling to room temperature, all the phantoms were stored in a refrigerated environment. They were allowed to warm at room temperature for two hours before scanning.

All phantoms were scanned using a 23-mm-diameter Tx/Rx 1H RF volume coil (Bruker BioSpin, Ettlingen, Germany). PETALUTE images were acquired with a total of 9,216 petals, each petal containing 206 sampling points with 103 sampling points per echo. RARE-VTR and PETALUTE $T_1$ maps were calculated as previously described. Regions of interest (ROIs) were defined in 3D Slicer[34] around the ferumoxytol. For each concentration, the mean and standard deviation of $T_1$ and $R_1$ were calculated within the ROI.

The relaxation rate $R_1$, defined as the reciprocal of $T_1$, has an approximately linear relationship with the concentration of ferumoxytol, as described in **Equation 7**[19].

$$R_1(C) = R_{1,0} + a \cdot C \qquad \text{Equation 7}$$

Where C is the ferumoxytol concentration (ppm), $R_{1,0}$ is the intrinsic $R_1$ of the agarose solution ($s^{-1}$), and a is the slope (ppm$^{-1}$ s$^{-1}$).

A calibration curve was established by linear regression analysis of $R_1$ and ferumoxytol concentration.

## In vivo Feasibility Study

Female BALB/c mice (8-10 weeks old) were obtained from Envigo (Lafayette, Indiana) and housed in the local facility under pathogen-free conditions. All experiment procedures were reviewed and approved by the local institution in accordance with ARRIVE guidelines.

4T1 cells were obtained from ATCC and cultured in RPMI-1640 medium containing 10% FBS and 1% penicillin/streptomycin. All cultures were incubated at 37°C under humidified conditions with 5% $CO_2$ and 95% air.

Three female BALB/c mice bearing 4T1 mammary and flank tumors (dual tumor model) were used in this experiment. To establish the model, mice were subcutaneously injected with a single-cell suspension of $1 \times 10^5$ 4T1 tumor cells in 50 μL of PBS into the fourth right mammary fat (designed as primary tumor), and $5 \times 10^4$ 4T1 cells in 50 μL of PBS into the left upper thigh (designed as flank tumor). Mice were divided into control (n = 1) and ferumoxytol-injected (n = 2) groups. Fourteen days after tumor implantation, the ferumoxytol group received a single dose of 40 mg/kg ferumoxytol administered via the tail vein.

MRI scans were conducted 24 hours after injection. All mice were scanned using a 7T MRI system with a 40-mm-diameter Tx/Rx 1H RF volume coil (Bruker BioSpin, Ettlingen, Germany), and anesthesia was maintained at 2% isoflurane in oxygen throughout the imaging process. PETALUTE images were acquired using 36,864 petals with 206 sampling points per petal (103 sampling points per echo). RARE-VTR images were also obtained for comparison.

ROIs were manually delineated in the thigh muscle (as a reference region), mammary tumors, and flank tumors using 3D Slicer[34], and quantitative analyses were performed in MATLAB. For each ROI, the mean and standard deviation of $T_1$ and $R_1$ were calculated from both PETALUTE and RARE-VTR $T_1$ maps to enable comparative evaluation.

## Result

### Phantom Study

**Figure 2** presents reconstructed images of all phantoms acquired using RARE-VTR and PETALUTE, demonstrating the positive contrast of ferumoxytol produced by the PETALUTE first echo and the negative contrast by the second echo. It should be noted that the PETALUTE images lost some positive contrast for ferumoxytol at a concentration of 5000 ppm. The RARE-VTR images did not have positive contrast at any tested ferumoxytol concentration.

To evaluate the quantitative sensitivity, the relationship between mean $R_1$ values and ferumoxytol concentration was analyzed (**Figure 3**). A significant positive linear correlation was observed for PETALUTE-based mean $R_1$ (R = 0.975, p < 0.01). In contrast, the RARE-VTR-based mean $R_1$ did not show significant correlation (R = 0.672, p = 0.144). Additionally, the PETALUTE $R_1$ standard deviations were lower than those for RARE-VTR, as shown in **Table 2**, indicating improved precision. The 5000 ppm phantom data were excluded from the linear regression model due to severe susceptibility artifacts in both PETALUTE and RARE-VTR images **(Figure 2).**

### In Vivo Feasibility Study

Reconstructed PETALUTE images cover the whole abdominal region and provide visualization of anatomical features **(Figure 4).** Notably, structures such as the kidneys and livers, which were poorly defined in the RARE-VTR images due to a limited field of view, were distinguishable using PETALUTE. In addition, while RARE-VTR images displayed a hypointense region within the flank tumor, PETALUTE showed slightly higher signal intensity near the tumor center **(Figure 5).**

ROIs within the thigh muscle, flank tumors, and mammary tumors allowed the determination of $T_1$ values from PETALUTE and RARE-VTR. The overall mean $T_1$ and $R_1$ values for each mouse and the respective ROI are reported in **Table 3**. For both $T_1$ mapping methods, muscle tissue mean $T_1$ values remained relatively stable across mice. For RARE-VTR, the mean muscle $T_1$ values for the ferumoxytol-injected mice were within 6% of the control mouse $T_1$ values. For PETALUTE, the mean muscle $T_1$ values for ferumoxytol-injected mice were within 4% of the control mouse $T_1$ values.

In the flank tumor, both $T_1$ mapping methods yielded lower mean $T_1$ values in ferumoxytol-injected mice than in control mouse. The $T_1$ reduction in the flank tumor in ferumoxytol-injected mice was approximately 10–12% with RARE-VTR and 4–7% with PETALUTE. In the mammary tumor, $T_1$ changes differed across the two $T_1$ mapping methods. With RARE-VTR, both the ferumoxytol-injected mice exhibited an approximately 9–11% $T_1$ reduction relative to the control mouse. PETALUTE, however, revealed divergent responses: relative to the control mouse (MS47), the mean $T_1$ increased by approximately 42% in MS48 but decreased by approximately 4% in MS49.

### Discussion

We developed a $B_1$-corrected PETALUTE-VFA $T_1$ mapping protocol that enables quantitative, high-resolution imaging of ferumoxytol. This new protocol improved the linearity between $R_1$ and ferumoxytol concentration, produced positive contrast at high ferumoxytol concentrations, and expanded spatial coverage, thereby overcoming the poor image quality and signal losses that limit conventional $T_1$ mapping for SPION imaging.

The improved quantitative performance of PETALUTE $T_1$ mapping in the phantom study is attributed to its ultrashort echo time, which enables signal acquisition from iron before substantial $T_2^*$-related dephasing occurs. This early acquisition results in higher SNR and reduced variability in estimated $T_1$ values. Consequently, the protocol delivers more precise quantitative measurements. It establishes a reliable calibration curve for $T_1$-based ferumoxytol quantification up to 1250 ppm, whereas the conventional RARE-VTR sequence failed to achieve comparable results. Furthermore, reduced susceptibility effects at lower magnetic field strengths are anticipated to extend the usable linear range of the calibration curve, suggesting that the limits reported here at 7T are likely conservative compared with low-field ferumoxytol quantification methods[35].

The in vivo scanning demonstrated the feasibility of using the PETALUTE $T_1$ mapping protocol to assess ferumoxytol distribution in rodent tumors. Given that no accumulation of ferumoxytol was expected in muscle tissue, the muscle served as a reference region. As expected, we observed minimal differences in $T_1$ values in this region between the control and ferumoxytol-injected groups, with mean muscle $T_1$ values within 6% for RARE-VTR and within 4% for PETALUTE relative to the control mouse. In tumors, accumulation of ferumoxytol was expected to lead to $T_1$ shortening, including decreased $T_1$ values and increased $T_1$-weighted MRI signal intensity. This behavior was observed in the flank tumors of ferumoxytol-injected mice in our study, which exhibited lower $T_1$ values with both methods and higher intratumoral signal intensity in the PETALUTE first echo. This finding supports the expected preferential accumulation of ferumoxytol within tumor tissue, consistent with prior reports of tumor-specific ferumoxytol uptake that manifests as decreased $T_1$ values[36]. The greater $T_1$ reduction observed with RARE-VTR is likely due to its longer TE (20 ms), which leads to substantial $T_2$ dephasing in the presence of ferumoxytol. This residual $T_2$ weighting reduces the effective MRI signal in RARE-VTR images, causing hypointensity and exaggerating the apparent $T_1$-shortening effect. In contrast, the ultrashort TE of PETALUTE (0.016 ms) minimizes $T_2^*$ decay, enabling acquisition of an accurate $T_1$-weighted signal before substantial dephasing occurs.

However, an unexpected result was observed in the in vivo study. One of the ferumoxytol-injected mice (MS48) exhibited a nearly 42% increase in mean PETALUTE $T_1$ in the mammary tumor. To investigate this unexpected finding, multimodal images were examined **(Figure 6)**. In the MS48 mammary tumor, the combination of $T_1$ hypointensity and persistent $T_2^*$ signal (indicated by dashed arrow in **Figure 6**) has been reported in previous studies as a marker of tumor necrosis[37]. These image features are also consistent with an increase in $T_1$ and $T_2^*$ values. Although originally reported in a different tumor model rather than in the 4T1 mammary tumor, increases in $T_1$ and $T_2^*$ are broadly associated with inflammation or necrosis[14]. Importantly, the larger tumor volume in the MS48 mammary tumor (MS47: 152 mm$^3$, MS48: 210 mm$^3$, MS49:189 mm$^3$) likely promoted necrosis, which dominated the MR signal response[37]. Based on the phantom study (**Figure 2**), ferumoxytol accumulation is anticipated to produce hyperintensity on the PETALUTE first echo and hypointensity on the second echo. Therefore, the hyperintensity region observed in the PETALUTE second echo of MS48 (**Figure 6**) is not consistent with ferumoxytol accumulation and is more plausibly attributed to a long $T_2$ component, such as fluid-like necrosis. Overall, these observations suggest that the hyperintensity observed in the MS48 tumor across both PETALUTE echoes likely reflects a mixture of necrotic and iron-rich tissue. The resulting PETALUTE $T_1$ map revealed elevated $T_1$ values in these potential necrosis-dominated regions. Interestingly, the RARE-VTR $T_1$ map was not sensitive to these effects **(Figure 6).** Similar intratumoral hyperintensity was also observed in both PETALUTE echoes of MS49, though without measurable $T_1$ elevation. This might arise from mild or early-stage necrosis, which was insufficient to alter the overall MR signal behavior. Taken together, these findings highlight PETALUTE's ability to capture biological information that may be missed with conventional sequences. Tumor necrosis is a well-established prognostic indicator, and a noninvasive means to measure it would be valuable. Further histological studies are required to confirm this interpretation.

This study has several limitations that should be acknowledged. First, PETALUTE images appear slightly blurrier than those acquired with RARE-VTR **(Figure 5).** This is primarily because we deliberately use a high undersampling factor together with a CS-SENSE reconstruction to enable faster, non-gated whole abdominal T1 mapping. For the target nominal resolution (matrix = 256 × 256 × 256, r = 128), the Nyquist criterion corresponds to approximately $4\pi \times r^2$ petals ($4\pi * 128^2 \approx 2.1 \times 10^5$). In comparison, our protocol uses 36,846, which is 18% of the Nyquist sampling density. This level of undersampling reduces image sharpness and fine edge detail, leading to some loss of anatomic definition relative to RARE-VTR. However, previous work has shown that PETALUTE $T_1$ estimates remain stable even at higher undersampling level (18,192 petals), so substantial degradation of $T_1$ values is not expected at the sampling density used here[38]. If higher anatomical clarity is desired, the protocol can be adapted by increasing the sampling density. Second, although the PETALUTE $T_1$ mapping approach extends the reliable quantitative range of ferumoxytol concentrations, it remains limited at very high concentrations. At 5000 ppm (5 mg/mL), the susceptibility of ferumoxytol at 7T results in a local frequency shift of approximately 20 kHz and an estimated $T_2^*$ of about 0.01 ms[39]. These large off-resonance effects and extremely short $T_2^*$ lead to considerable signal attenuation and k-space misregistration that severely degrade $T_1$ mapping accuracy.

Future studies will focus on validating in vivo findings with histological evidence. Further evaluation of PETALUTE's performance will also be carried out using simulations to understand how varying SPION concentration, off-resonance, and $T_2^*$ shortening affect $T_1$ mapping precision.

## Conclusion

In summary, the PETALUTE $T_1$ mapping protocol provides more reliable ferumoxytol quantification compared with RARE-VTR. Uniquely, this protocol enables high-resolution, non-gated whole-abdominal images and dual-echo imaging within a single scan. This multimodal capability makes PETALUTE particularly valuable for monitoring treatment with SPION-based therapies.

## Acknowledgments

This work was supported by NIH grant 1R01NS131160. Additional support from Eli Lilly and Company is gratefully acknowledged. The authors also gratefully acknowledge the support from the Institute for Cancer Research, NIH grant P30 CA023168.

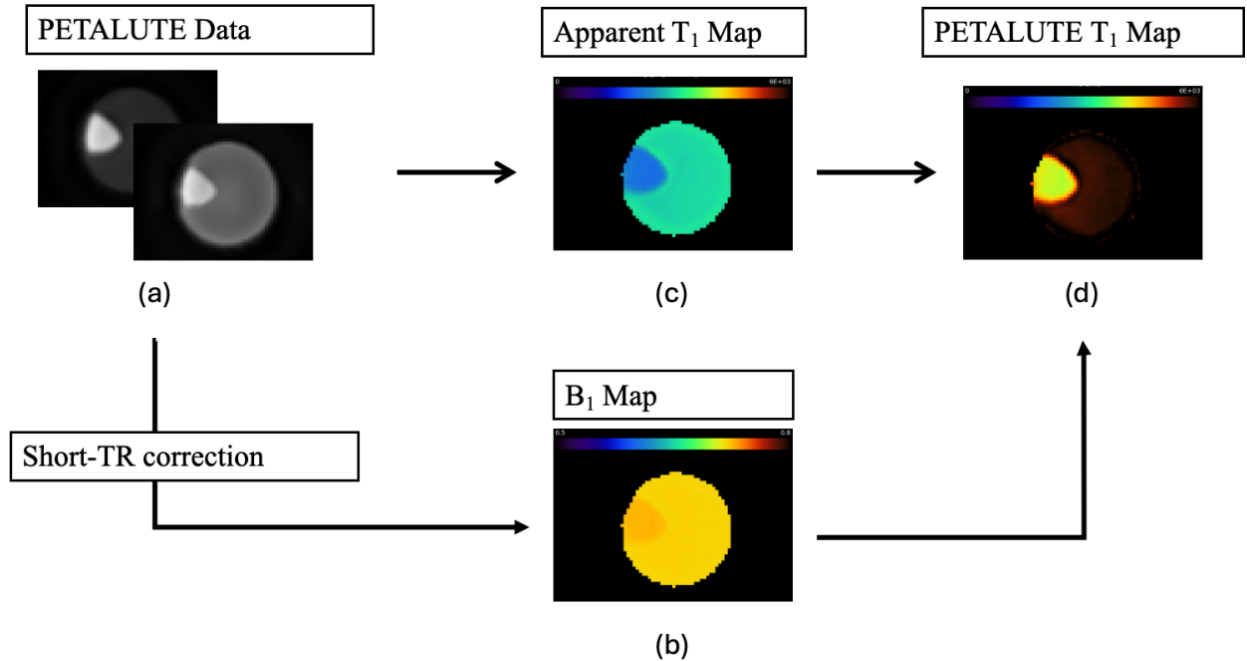

**Figure 1.** (a) PETALUTE image reconstructed using nonuniform fast Fourier transform method. (b) $B_1$ map generated by the variable flip angle method with short-TR correction and smoothed using a Gaussian filter with σ = 3; color bar range from 0.5 to 0.8. (c) Apparent $T_1$ map generated using the two flip-angle PETALUTE images; color bar range from 0 to 6000 ms. (d) PETALUTE $T_1$ map derived from the apparent $T_1$ map corrected using the $B_1$ map; color bar range from 0 to 6000 ms.

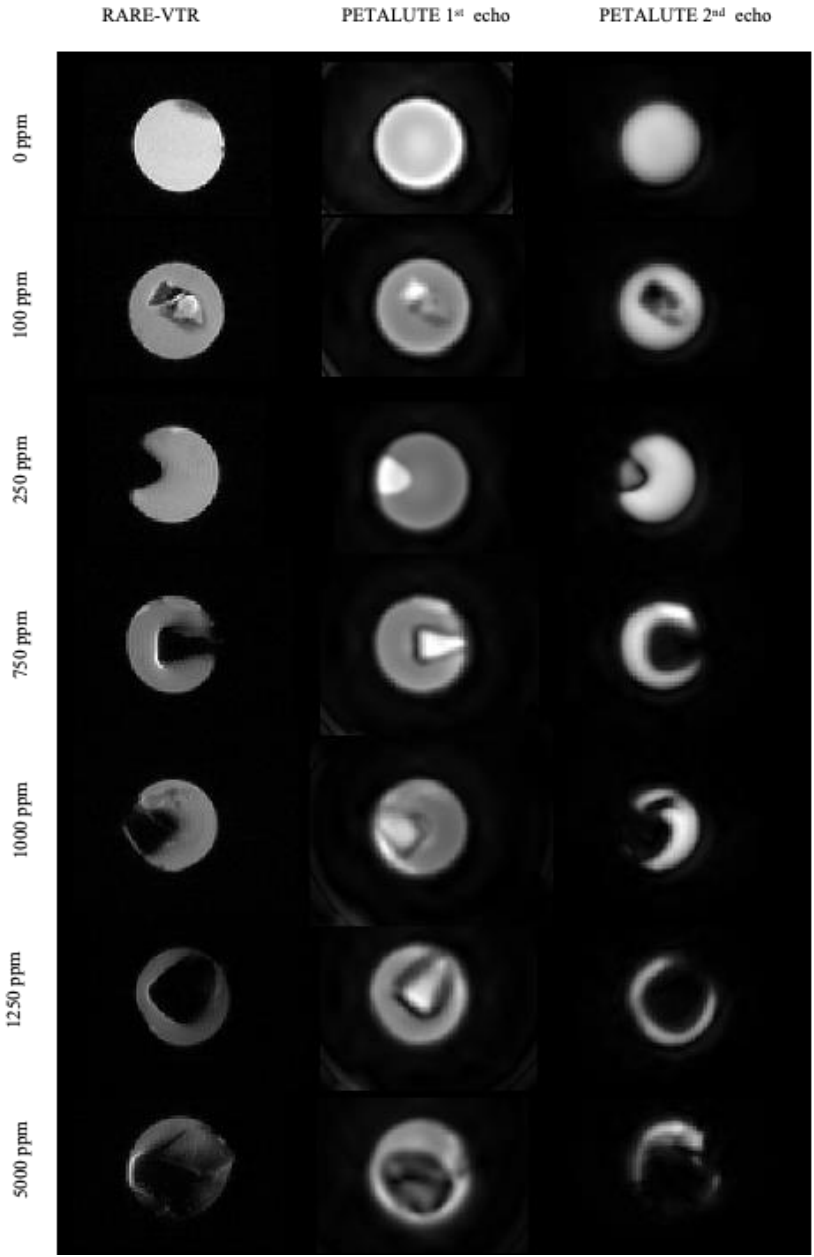

**Figure 2.** Reconstructed images of phantoms acquired using RARE-VTR and PETALUTE.
Phantom phantoms containing various concentrations of ferumoxytol (0 – 5000 ppm) appear as bright regions in PETALUTE first echo (flip angle = 4) and dark regions in PETALUTE second echo (flip angle = 4) and RARE-VTR images (TR = 934.58 ms).

Table 1. Image protocol parameters

|  | RARE-VTR | MGE | PETALUTE |
|---|---|---|---|
| Resolution | 0.234 × 0.234 × 1 mm³ | 0.234 × 0.234 × 1 mm³ | 0.2 × 0.2 × 0.2 mm³ |
| FOV | 30 × 30 mm³<br>20 slices (thickness = 1 mm) | 30 × 30 mm³<br>20 slices (thickness = 1 mm) | 40 × 40 × 40 mm³ (Phantom)<br>60 × 60 × 60 mm³ (In-vivo) |
| TE | 20.3 ms | 2 ms, 6 ms, 10 ms, 14 ms,<br>18 ms, 22 ms, 26 ms, 30 ms,<br>34 ms, 38 ms | 0.016 ms (first echo)<br>0.032 ms (second echo) |
| TR | 934.58 ms, 1788.44 ms,<br>2986.55 ms, 5011.78 ms,<br>15000.0 ms | 1500 ms | 7 ms |
| Flip Angle | NA |  | 4°, 20° |
| Acquisition Time | 6.18 min | 6.4 min | 1.08 min/flip-angle (Phantom)<br>4.31 min/flip-angle (In-vivo) |

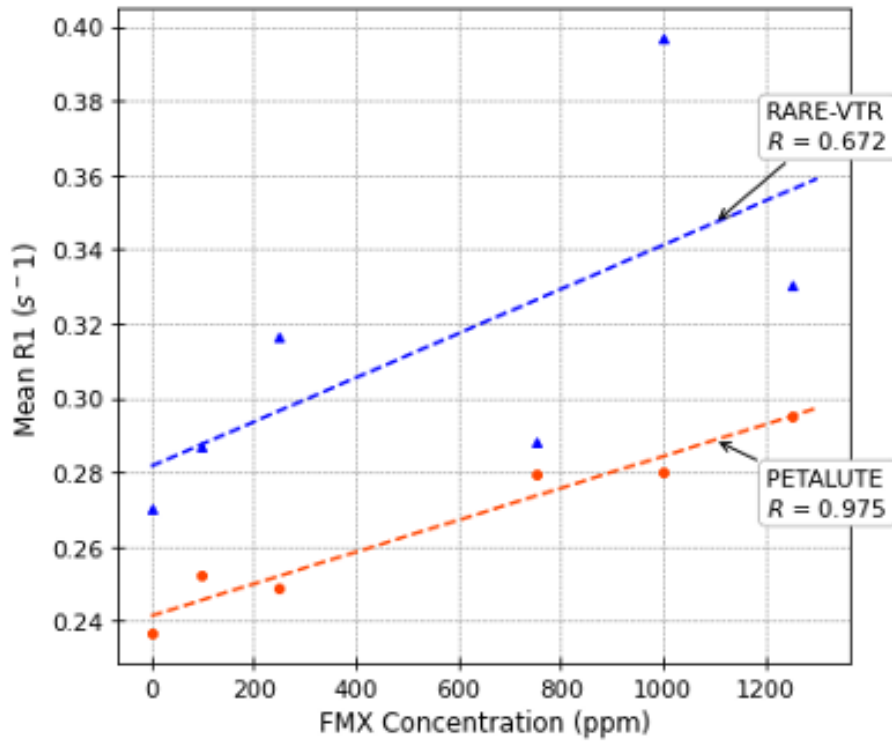

**Figure 3.** Linear regression of mean $R_1$ versus ferumoxytol (FMX) concentration for PETALUTE (circles) and RARE-VTR (triangles). The R values were 0.975 for PETALUTE and 0.672 for RARE-VTR. Data at 5000 ppm were excluded as outliers due to extreme artifacts in both methods.

**Table 2.** Summary of $T_1$ and $R_1$ for each ferumoxytol (FMX) phantom measured using RARE-VTR and PETALUTE methods. Data are presented as mean ± standard deviation (SD).

| FMX Concentration (ppm) | RARE-VTR $R_1$ Mean ± SD ($s^{-1}$) | PETALUTE $R_1$ Mean ± SD ($s^{-1}$) | RARE-VTR $T_1$ Mean ± SD (ms) | PETALUTE $T_1$ Mean ± SD (ms) |
|---|---|---|---|---|
| 0 | 0.270 ± 0.008 | 0.237 ± 0.002 | 3703 ± 103 | 4229 ± 41 |
| 100 | 0.287 ± 0.038 | 0.252 ± 0.014 | 3534 ± 427 | 3978 ± 237 |
| 250 | 0.316 ± 0.104 | 0.249 ± 0.002 | 4170 ± 3874 | 4008 ± 31 |
| 750 | 0.288 ± 0.126 | 0.279 ± 0.004 | 4126 ± 1782 | 3581 ± 49 |
| 1000 | 0.397 ± 0.184 | 0.280 ± 0.005 | 2876 ± 949 | 3574 ± 58 |
| 1250 | 0.331 ± 0.255 | 0.295 ± 0.006 | 5166 ± 4659 | 3387 ± 69 |
| 5000 | 0.826 ± 0.448 | 0.270 ± 0.011 | 1639 ± 983 | 3706 ± 160 |

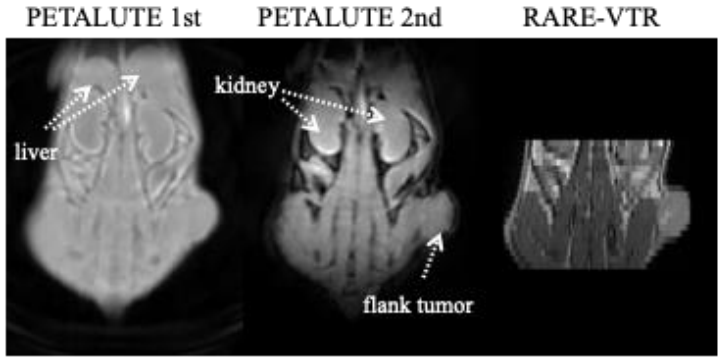

**Figure 4.** Exemplary coronal images from a control mouse (MS47) acquired using PETALUTE first echo and second echo (FA = 4°) and RARE-VTR (TR = 934.58 ms).

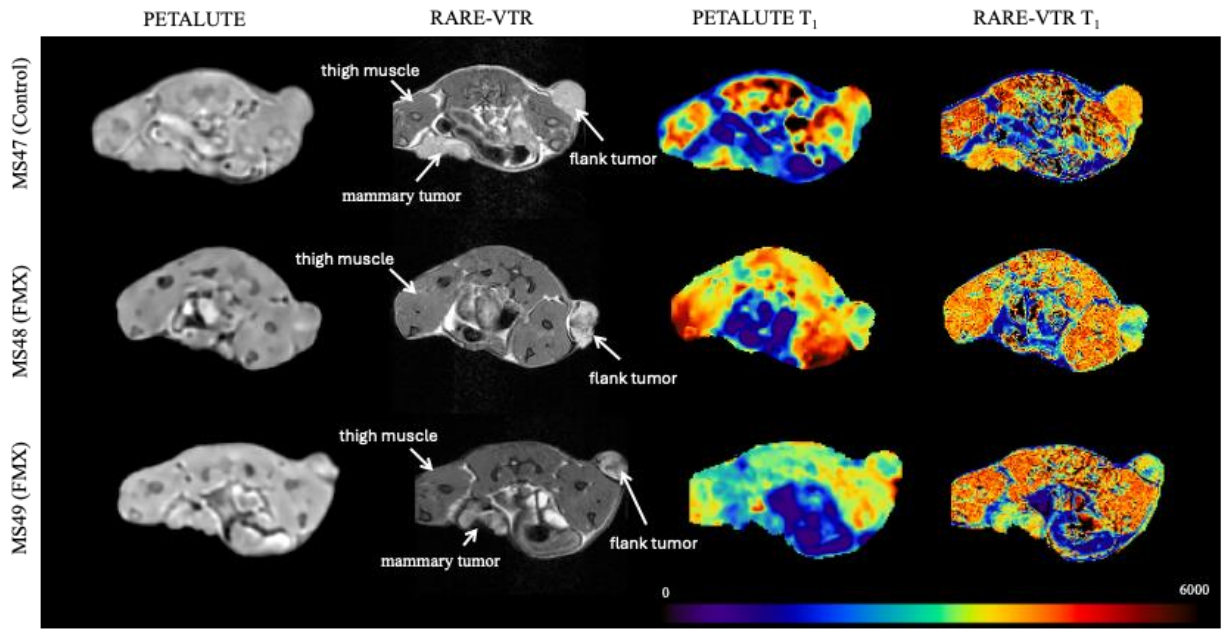

**Figure 5.** PETALUTE and RARE-VTR images and corresponding $T_1$ maps from the control mouse (MS47) and the mice with ferumoxytol (FMX) injection (MS48 and MS49). PETALUTE images are first echo acquired with a flip angle of 4°, while RARE-VTR images are acquired with TR = 934.58 ms. Thigh muscle, flank tumors and mammary tumors are indicated (the mammary tumor of MS48 appears in a different slice and is not shown here). Color maps show $T_1$ values in milliseconds, with a scale range of 0–6000 ms.

**Table 3.** Mean and standard deviation (SD) of $T_1$ and $R_1$ values within the ROIs of thigh muscle, flank tumors and mammary tumors in three mice.

| ID | Group | ROI | RARE-VTR $R_1$ Mean ± SD (s⁻¹) | PETALUTE $R_1$ Mean ± SD (s⁻¹) | RARE-VTR $T_1$ Mean ± SD (ms) | PETALUTE $T_1$ Mean ± SD (ms) |
|---|---|---|---|---|---|---|
| MS 47 | Control | Thigh Muscle | 0.242 ± 0.023 | 0.247 ± 0.015 | 4164 ± 392 | 4063 ± 249 |
| | | Flank Tumor | 0.260 ± 0.012 | 0.298 ± 0.043 | 3850 ± 184 | 3424 ± 464 |
| | | Mammary Tumor | 0.265 ± 0.029 | 0.374 ± 0.063 | 3821 ± 396 | 2745 ± 443 |
| MS 48 | Ferumoxytol | Thigh Muscle | 0.258 ± 0.029 | 0.240 ± 0.021 | 3919 ± 427 | 4203 ± 357 |
| | | Flank Tumor | 0.295 ± 0.037 | 0.318 ± 0.042 | 3438 ± 389 | 3198 ± 416 |
| | | Mammary Tumor | 0.302 ± 0.073 | 0.265 ± 0.048 | 3472 ± 691 | 3894 ± 667 |
| MS 49 | Ferumoxytol | Thigh Muscle | 0.239 ± 0.029 | 0.251 ± 0.020 | 4252 ± 545 | 4002 ± 322 |
| | | Flank Tumor | 0.306 ± 0.060 | 0.307 ± 0.030 | 3369 ± 556 | 3286 ± 341 |
| | | Mammary Tumor | 0.302 ± 0.060 | 0.386 ± 0.049 | 3401 ± 500 | 2632 ± 330 |

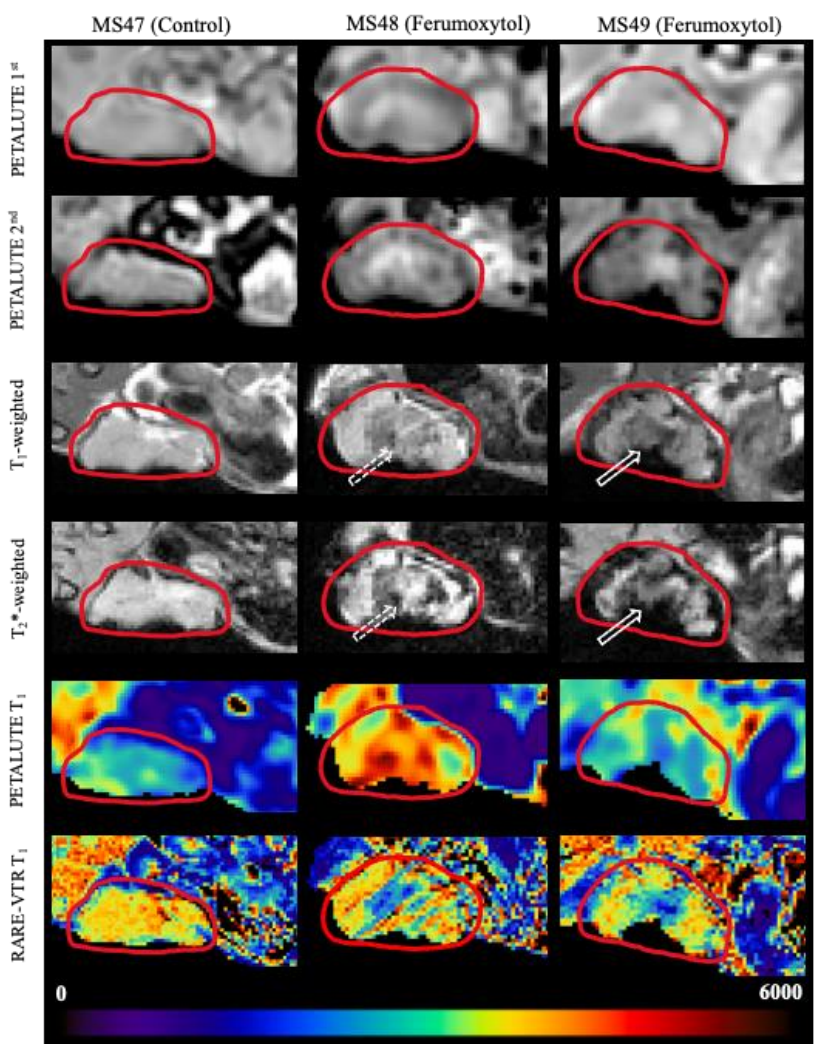

**Figure 6.** Multimodal images of three included animals focusing on the mammary tumor. Shown are PETALUTE first echo, PETALUTE second echo, $T_1$-weighted (RARE-VTR, TE = 20 ms, TR = 933 ms), $T_2$*-weighted (MGE, TE = 6 ms, TR = 1500 ms), PETALUTE $T_1$ map, and RARE-VTR $T_1$ map images. Mammary tumors are contoured in red. Dashed arrows indicate $T_1$ hypointensity with persistent $T_2$* signal indicated in MS48 while solid arrows indicate $T_1$ hypointensity with $T_2$* signal loss in MS49. MS47 (control) showed uniform intratumor intensity. Color maps display $T_1$ values in milliseconds, with a scale range of 0–6000 ms.